# Energy Efficiency in Secure Multi-Antenna Systems

Alessio Zappone, Pin-Hsun Lin, and Eduard A. Jorswieck
Communications Laboratory, TU Dresden, Germany
Email: {Alessio.Zappone, Pin-Hsun.Lin, Eduard.Jorswieck}@tu-dresden.de

***Abstract*—The problem of resource allocation in multiple-antenna wiretap channels is investigated, wherein a malicious user tries to eavesdrop the communication between two legitimate users. Both multiple input single output single-antenna eavesdropper (MISO-SE) and multiple input multiple output multiple-antenna eavesdropper (MIMO-ME) systems are considered. Unlike most papers dealing with physical layer security, the focus of the resource allocation process here is not to maximize the secrecy capacity, but rather to maximize the energy efficiency of the system. Two fractional energy-efficient metrics are introduced, namely the ratios between the system secrecy capacity and the consumed power, and between the system secret-key rate and the consumed power. Both performance metrics are measured in bit/Joule, and result in non-concave fractional optimization problems, which are tackled by fractional programming theory and sequential convex optimization. For both performance metrics, the energy-efficient resource allocation is carried out considering both perfect as well as statistical channel state information (CSI) as to the channel from the legitimate transmitter to the eavesdropper.**

## I. Introduction

Wireless communications are inherently subject to attacks from malicious users, due to their broadcast nature. Since the seminal paper [1], physical layer security has studied the transmission techniques to achieve unconditionally confidential communication in presence of eavesdroppers [2], [3]. Unlike traditional cryptography which requires private key generation and distribution, physical layer security exploits the nature of the communication channel to confuse the eavesdropper, transmitting at a lower rate than the system secrecy capacity. However, secrecy capacity optimization is usually not energy-efficient, because it typically requires to transmit at the maximum feasible power. Indeed, in MISO-SE systems with perfect CSI of the legitimate channel and statistical CSI of the eavesdropper's channel, [4] proves that, under mild conditions, using full transmit power is optimal. As for MIMO-ME systems with statistical CSI, full transmit power is shown to be optimal in [5, Theorem 1], whereas with perfect CSI, [6] shows that full transmit power is optimal for degraded wiretap channels.

As a result, secrecy capacity maximization leads to large energy consumptions, which are not desirable in present and forthcoming communication networks. Information and communication technologies (ICT) are responsible for about 5% of the entire world energy consumption, and this percentage is rapidly growing due to the exponential increase of connected devices and infrastructure nodes. It is foreseen that the number of connected nodes will reach 50 billion by 2020 and that the energy demand will soon become unmanageable [7]. Moreover, the resulting greenhouse gas emissions and electromagnetic pollution will exceed safety thresholds. For these reasons, energy efficiency is rapidly becoming a critical performance measure for future wireless networks, which, in addition, also require a massive increase of secure and safe communications.

These remarks motivate us to study the problem of resource allocation for secure communications from an energy-efficient perspective. Rather than the well-investigated secrecy capacity optimization problem [8]–[13], energy-efficient performance measures should be introduced, which are able to trade-off between the contrasting needs to ensure a reliable and secure communication, and to limit power consumptions. From a physical standpoint, the energy efficiency of a communication system is defined as the benefit-cost ratio in terms of amount of information reliably transmitted and of consumed energy. In systems which do not require a secret communication, this leads to the well-established definition of the energy efficiency as the ratio between the system capacity (or achievable rate) over the consumed power [14]. The natural extension to the case in which a reliable and *secure* communication is required, is to consider the ratio between the secrecy capacity (or secrecy rate) and the consumed power. This metric, which we label secrecy energy efficiency (SEE), is measured in bit/Joule and represents the amount of bits which can be reliably and secretly transmitted per Joule of consumed energy.

The SEE is a natural measure to consider in scenarios where secrecy is ensured by transmitting at a lower data-rate than the system secrecy capacity. However, as already mentioned, a more traditional technique to ensure secrecy is to rely on a cryptographic key. In this case, physical layer security tools can still prove very useful to allow secure key generation and the concept of secret-key rate has been introduced [3], which measures the amount of bits securely distributed over a public channel between the legitimate users for key-generation purposes [15]. Pioneering works on secret key agreement using physical layer security arguments are [16]–[18], which considered a broadcast channel with a public channel for communications between the users, while in [19], [20] key sharing was considered in fading channels. In systems using a cryptographic key for secrecy, a more suitable definition of energy efficiency appears to be the ratio between the system secret-key rate, and the consumed power, which we label as Secret-key energy efficiency (SKEE). As the SEE, the SKEE is measured in bit/Joule, and it represents the number of bits which can be securely distributed over a public channel per Joule of consumed energy.

Very few previous contributions on energy efficiency in wiretap channels have appeared, which however consider neither the SEE nor the SKEE. We mention [21] where the ratio between the system outage probability and the consumed energy is optimized for a downlink orthogonal frequency division multiple access (OFDMA) network, [22] where the secrecy-energy trade-off in single-antenna, single-carrier Gaussian wiretap channels is studied in terms of actual communication rate and consumed energy, and [23], which provides an information-theoretic analysis of secret communications per transmission cost.

Motivated by this background, unlike previous contributions, in this work we study the problem of energy-efficient resource allocation in multiple-antenna systems. Besides introducing the two new energy-efficient metrics of SEE and SKEE, we make the following contributions:

- we consider the maximization of the SEE and of the SKEE both in MISO-SE and MIMO-ME systems, which leads to challenging non-concave, fractional problems. As for MISO-SE systems, we leverage generalized concavity theory to provide transmit power and beamforming vector optimization algorithms, which are able to globally maximize the considered energy-efficient metrics, with limited complexity. As for the MIMO-ME case, the optimization problems are tackled by fractional programming theory as far as SKEE maximization is concerned and by combining fractional programming theory with the framework of sequential convex optimization, for SEE maximization. The proposed approach globally maximizes the SKEE, and is guaranteed to converge to a point fulfilling the Karush Kuhn Tucker (KKT) optimality conditions of the SEE maximization problem;
- we develop the described resource allocation methods considering both the scenarios in which the legitimate transmitter has perfect CSI of all propagation channels, and that in which the legitimate transmitter only knows the statistics of the channel to the eavesdropper;
- the proposed energy-efficient algorithms can be readily specialized to the case in which the secrecy rate or the secret-key rate are to be optimized. In the general MIMO-ME scenario, secrecy rate optimization is still an open problem, and numerical results suggest that our approach outperforms competing alternatives and achieves near-optimal performance.

The remainder of the paper is organized as follows. Section II describes the system model and formulates the considered optimization problems. Sections III and IV contain the main contributions of the paper, providing the energy-efficient resource allocation algorithms for the MISO-SE and MIMO-ME scenarios, respectively, with both perfect and statistical CSI. Section V illustrates numerical results to assess the performance of the proposed resource allocation methods. Finally, Section VI contains concluding remarks. Important basic results on generalized concavity and fractional programming is provided in Appendix A.

## II. System model

The considered system model is depicted in Fig. 1. A legitimate transmitter (Alice) is equipped with $N_A$ antennas and wants to have a secure communication with its intended receiver (Bob), equipped with $N_B$ antennas, in presence of an eavesdropper (Eve), equipped with $N_E$ antennas. Denote by $\boldsymbol{H}$ and $\boldsymbol{G}$ the $N_B \times N_A$ and $N_E \times N_A$ matrix channels between Alice and Bob, and Alice and Eve, respectively, and by $\boldsymbol{Q}$ the transmit covariance matrix at Alice, subject to the power constraint $\mathrm{tr}(\boldsymbol{Q}) \leq P_{max}$, with $P_{max}$ the maximum feasible transmit powers[1]. Then, the total power consumed in the system is given by the sum of the transmit power $\mu\mathrm{tr}(\boldsymbol{Q})$, with $\mu = 1/\eta$ and $\eta$ the transmit amplifier efficiency, and the hardware power $P_c$ required to operate the legitimate system.

As mentioned in the introduction, two energy-efficient performance metrics will be considered.

- The SEE, defined as the ratio between the secrecy rate and the consumed power, namely

$$\mathrm{SEE} = \frac{\log\left(\frac{\left|\boldsymbol{I}_{N_B} + \boldsymbol{H}\boldsymbol{Q}\boldsymbol{H}^H\right|}{\left|\boldsymbol{I}_{N_E} + \boldsymbol{G}\boldsymbol{Q}\boldsymbol{G}^H\right|}\right)}{\mu\mathrm{tr}(\boldsymbol{Q}) + P_c}, \tag{1}$$

  which represents the amount of bits that can be reliably and secretly transmitted without being eavesdropped by Eve.
- The SKEE, defined as the ratio between the secret-key rate and the consumed power, namely

$$\mathrm{SKEE} = \frac{\log\left(\frac{\left|\boldsymbol{I}_{N_A} + \boldsymbol{Q}\left(\boldsymbol{H}^H\boldsymbol{H} + \boldsymbol{G}^H\boldsymbol{G}\right)\right|}{\log\left|\boldsymbol{I}_{N_A} + \boldsymbol{Q}\boldsymbol{G}^H\boldsymbol{G}\right|}\right)}{\mu\mathrm{tr}(\boldsymbol{Q}) + P_c}, \tag{2}$$

  which represents the amount of cryptographic-key bits per Joule of consumed energy, on which the legitimate users can secretly agree. For the derivation of the secret-key rate at the numerator of (2) we refer to [24, Lemma 2].

Assuming Alice has perfect CSI of both channels $\boldsymbol{H}$ and $\boldsymbol{G}$, (1) and (2) can be directly used for resource allocation purposes. However, in some practical instances of communication systems, it is more realistic to assume that Alice does not perfectly known the channel to the eavesdropper, but only its statistics[2]. In this case, the ergodic counterparts of (1) and (2)

[1] In some scenarios it might be useful to set a positive minimum transmit power $P_{min}$, for quality-of-service reasons. The results to be presented can be straightforwardly extended to such a scenario.

[2] In any case we assume that Bob has perfect CSI of $\boldsymbol{H}$, and Eve has perfect CSI of $\boldsymbol{H}$ and $\boldsymbol{G}$.

should be considered for resource allocation purposes, namely

$$\text{SEE}^{\text{erg}} = \frac{\mathbb{E}_{\boldsymbol{G}}\left[\log\left(\frac{\left|\boldsymbol{I}_{N_B}+\boldsymbol{H}\boldsymbol{Q}\boldsymbol{H}^H\right|}{\left|\boldsymbol{I}_{N_E}+\boldsymbol{G}\boldsymbol{Q}\boldsymbol{G}^H\right|}\right)\right]}{\mu\text{tr}(\boldsymbol{Q})+P_c} \quad (3)$$

$$\text{SKEE}^{\text{erg}} = \frac{\mathbb{E}_{\boldsymbol{G}}\left[\log\left(\frac{\left|\boldsymbol{I}_{N_A}+\boldsymbol{Q}\left(\boldsymbol{H}^H\boldsymbol{H}+\boldsymbol{G}^H\boldsymbol{G}\right)\right|}{\left|\boldsymbol{I}_{N_A}+\boldsymbol{Q}\boldsymbol{G}^H\boldsymbol{G}\right|}\right)\right]}{\mu\text{tr}(\boldsymbol{Q})+P_c} . \quad (4)$$

The problem to be tackled in the rest of this work is the maximization of both SEE and $\text{EE}_{key}$, with respect to the choice of the legitimate user's covariance matrix $\boldsymbol{Q}$, considering both the perfect and statistical CSI cases. In the next two sections, the problem will be first analyzed for the special case in which Bob and Eve are equipped with a single antenna, and then in the more challenging scenario in which multiple antennas are equipped at all nodes. In the former case, we will show that the problem can be reformulated as a fractional problem whose objective has a concave numerator and an affine denominator. The theory of generalized concavity, which is briefly reviewed in Appendix A, ensures that such an objective is a pseudo-concave (PC) function, for which first-order optimality conditions are necessary and sufficient. Instead, the latter scenario is more challenging and results in vector-valued or matrix-valued fractional problems, which are tackled leveraging fractional programming theory. The tools and results of fractional programming theory to be used in the sequel are also reviewed in Appendix A.

## III. Resource allocation for SEE maximization

When $N_B = N_E = 1$, let us denote by $\boldsymbol{w}$ and $p$ the legitimate user's unit-norm beamforming vector and transmit power, respectively, subject to the constraint $p \leq P_{max}$. Then, the expressions of (1) and (2) simplify to

$$\text{SEE} = \frac{\log\left(\frac{1+p\boldsymbol{w}^H\boldsymbol{h}\boldsymbol{h}^H\boldsymbol{w}}{1+p\boldsymbol{w}^H\boldsymbol{g}\boldsymbol{g}^H\boldsymbol{w}}\right)}{\mu p+P_c} , \quad (5)$$

and

$$\text{SKEE} = \frac{\log\left(\frac{1+p\boldsymbol{w}^H(\boldsymbol{h}\boldsymbol{h}^H+\boldsymbol{g}\boldsymbol{g}^H)\boldsymbol{w}}{1+p\boldsymbol{w}^H\boldsymbol{g}\boldsymbol{g}^H\boldsymbol{w}}\right)}{\mu p+P_c} , \quad (6)$$

respectively. The maximization of (5) and (6) are dealt with in Sections III-A and III-B, respectively, for the perfect CSI case, while the case of statistical CSI is addressed in Sections III-C and III-D.

### A. SEE maximization with perfect CSI

If Alice has perfect knowledge of the channel $\boldsymbol{g}$, the SEE maximization problem can be formulated as the optimization program

$$\max_{p,\boldsymbol{w}} \ \text{SEE} \quad (7a)$$
$$\text{s.t. } p \in [0, P_{max}] \quad (7b)$$
$$\|\boldsymbol{w}\| = 1 . \quad (7c)$$

To begin with, we see that the beamforming vector $\boldsymbol{w}$ affects only the numerator of (7a) and therefore should be chosen as the unit-norm vector that maximizes the secrecy rate. It is known that the solution to this problem is the dominant, unit-norm, generalized eigenvector of the matrix pencil $(\boldsymbol{I}_{N_A} + p\boldsymbol{h}\boldsymbol{h}^H \ , \ \boldsymbol{I}_{N_A} + p\boldsymbol{g}\boldsymbol{g}^H)$. Plugging the optimal $\boldsymbol{w}$ into (7), we obtain the power control problem

$$\max_{p,\boldsymbol{w}} \ \frac{\log\left(\lambda_{gen}(p)\right)}{\mu p+P_c} \quad (8a)$$
$$\text{s.t. } p \in [0, P_{max}] , \quad (8b)$$

wherein $\lambda_{gen}(p)$ denotes the maximum generalized eigenvalue of $(\boldsymbol{I}_{N_A} + p\boldsymbol{h}\boldsymbol{h}^H \ , \ \boldsymbol{I}_{N_A} + p\boldsymbol{g}\boldsymbol{g}^H)$. The dependence of $\lambda_{gen}$ on $p$ is rather involved, but it can be expressed in closed-form, leveraging the following lemma.

**Lemma 1.** *Let $\boldsymbol{x}$ and $\boldsymbol{y}$ be two arbitrary $N$-dimensional vectors. The matrix $\boldsymbol{x}\boldsymbol{x}^H - \boldsymbol{y}\boldsymbol{y}^H$ has at most two non-zero eigenvalues $(\lambda_1, \lambda_2)$ which are expressed as in (9) on top of next page.*

This result is stated without a proof in [8]. A proof can be given as follows.

*Proof:* The matrix $\boldsymbol{x}\boldsymbol{x}^H - \boldsymbol{y}\boldsymbol{y}^H$ has at most rank 2 and therefore has at most two non-zero eigenvalues, $\lambda_1$ and $\lambda_2$. Then we have,

$$\begin{aligned} \lambda_1 + \lambda_2 &= \text{tr}(\boldsymbol{x}\boldsymbol{x}^H - \boldsymbol{y}\boldsymbol{y}^H) = \|\boldsymbol{x}\|^2 - \|\boldsymbol{y}\|^2 , \\ (1+\lambda_1)(1+\lambda_2) &= \left|\boldsymbol{I}_N + \boldsymbol{x}\boldsymbol{x}^H - \boldsymbol{y}\boldsymbol{y}^H\right| . \end{aligned} \quad (10)$$

Elaborating on the second equation in (10), we obtain

$$\begin{aligned} &\left|\boldsymbol{I}_N + \boldsymbol{x}\boldsymbol{x}^H - \boldsymbol{y}\boldsymbol{y}^H\right| \\ &= \left|\boldsymbol{I}_N + \boldsymbol{x}\boldsymbol{x}^H\right|\left|\boldsymbol{I}_N - (\boldsymbol{I}_N + \boldsymbol{x}\boldsymbol{x}^H)^{-1}\boldsymbol{y}\boldsymbol{y}^H\right| \\ &= (1+\|\boldsymbol{x}\|^2)\left(1 - \boldsymbol{y}^H\left(\boldsymbol{I}_N - \frac{\boldsymbol{x}\boldsymbol{x}^H}{1+\|\boldsymbol{x}\|^2}\right)\boldsymbol{y}\right) \\ &= 1 + \|\boldsymbol{x}\|^2 - \|\boldsymbol{y}\|^2 - \|\boldsymbol{x}\|^2\|\boldsymbol{y}\|^2 + |\boldsymbol{y}^H\boldsymbol{x}|^2 . \end{aligned} \quad (11)$$

Then, solving for $\lambda_1$ in (10) we obtain

$$\begin{aligned} &\lambda_1^2 - (\|\boldsymbol{x}\|^2 - \|\boldsymbol{y}\|^2)\lambda_1 + |\boldsymbol{y}^H\boldsymbol{x}|^2 - \|\boldsymbol{x}\|^2\|\boldsymbol{y}\|^2 = 0 , \\ &\lambda_2 = \|\boldsymbol{x}\|^2 - \|\boldsymbol{y}\|^2 - \lambda_1 . \end{aligned} \quad (12)$$

Solving (12) yields (9). ■

This result can be applied to our scenario as follows. Observe that the maximum generalized eigenvalue of $(\boldsymbol{I}_{N_A} + p\boldsymbol{h}\boldsymbol{h}^H, \boldsymbol{I}_{N_A} + p\boldsymbol{g}\boldsymbol{g}^H)$ is equal to the maximum eigenvalue of the matrix $\boldsymbol{Z}$ shown in (13) at the top of next page, wherein we have used the matrix inversion lemma and we have denoted by $\boldsymbol{U}\boldsymbol{\Lambda}^{1/2}\boldsymbol{U}^H$ the eigenvalue decomposition (EVD) of $\left(\boldsymbol{I}_{N_A} - p\frac{\boldsymbol{g}\boldsymbol{g}^H}{1+p\|\boldsymbol{g}\|^2}\right)^{1/2}$. In particular, $\boldsymbol{U} = [\boldsymbol{u}_1, \ldots, \boldsymbol{u}_{N_A}]$ is such that $\boldsymbol{u}_1 = \boldsymbol{g}/\|\boldsymbol{g}\|$, while the other $N_A - 1$ columns are an orthonormal basis of the orthogonal complement of $\boldsymbol{u}_1$.

$$\lambda_1 = \frac{1}{2}\left(\|\boldsymbol{x}\|^2 - \|\boldsymbol{y}\|^2 + \sqrt{(\|\boldsymbol{x}\|^2 - \|\boldsymbol{y}\|^2)^2 - 4(|\boldsymbol{y}^H\boldsymbol{x}|^2 - \|\boldsymbol{x}\|^2\|\boldsymbol{y}\|^2)}\right)$$
$$\lambda_2 = \frac{1}{2}\left(\|\boldsymbol{x}\|^2 - \|\boldsymbol{y}\|^2 - \sqrt{(\|\boldsymbol{x}\|^2 - \|\boldsymbol{y}\|^2)^2 - 4(|\boldsymbol{y}^H\boldsymbol{x}|^2 - \|\boldsymbol{x}\|^2\|\boldsymbol{y}\|^2)}\right) \tag{9}$$

$$\begin{aligned}\boldsymbol{Z} &= \left(\boldsymbol{I}_{N_A} + p\boldsymbol{g}\boldsymbol{g}^H\right)^{-1/2}\left(\boldsymbol{I}_{N_A} + p\boldsymbol{h}\boldsymbol{h}^H\right)\left(\boldsymbol{I}_{N_A} + p\boldsymbol{g}\boldsymbol{g}^H\right)^{-1/2} \\ &= \left(\boldsymbol{I}_{N_A} + p\boldsymbol{g}\boldsymbol{g}^H\right)^{-1} + p\left(\boldsymbol{I}_{N_A} + p\boldsymbol{g}\boldsymbol{g}^H\right)^{-1/2}\boldsymbol{h}\boldsymbol{h}^H\left(\boldsymbol{I}_{N_A} + p\boldsymbol{g}\boldsymbol{g}^H\right)^{-1/2} \\ &= \boldsymbol{I}_{N_A} - p\frac{\boldsymbol{g}\boldsymbol{g}^H}{1+p\|\boldsymbol{g}\|^2} + p\left(\boldsymbol{I}_{N_A} - p\frac{\boldsymbol{g}\boldsymbol{g}^H}{1+p\|\boldsymbol{g}\|^2}\right)^{1/2}\boldsymbol{h}\boldsymbol{h}^H\left(\boldsymbol{I}_{N_A} - p\frac{\boldsymbol{g}\boldsymbol{g}^H}{1+p\|\boldsymbol{g}\|^2}\right)^{1/2} \\ &= \boldsymbol{I}_{N_A} - p\frac{\boldsymbol{g}\boldsymbol{g}^H}{1+p\|\boldsymbol{g}\|^2} + p\boldsymbol{U}\boldsymbol{\Lambda}^{1/2}\boldsymbol{U}^H\boldsymbol{h}\boldsymbol{h}^H\boldsymbol{U}\boldsymbol{\Lambda}^{1/2}\boldsymbol{U}^H\end{aligned} \tag{13}$$

$$\begin{aligned}\boldsymbol{Q} &= p\boldsymbol{U}\boldsymbol{\Lambda}^{1/2}\boldsymbol{U}^H\boldsymbol{h}\boldsymbol{h}^H\boldsymbol{U}\boldsymbol{\Lambda}^{1/2}\boldsymbol{U}^H - p\frac{\boldsymbol{g}\boldsymbol{g}^H}{1+p\|\boldsymbol{g}\|^2} = \\ &\left(\sqrt{\frac{p}{1+p\|\boldsymbol{g}\|^2}}(\boldsymbol{u}_1^H\boldsymbol{h})\boldsymbol{u}_1 + \sqrt{p}\sum_{i=2}^{N_A}(\boldsymbol{u}_i^H\boldsymbol{h})\boldsymbol{u}_i\right)\left(\sqrt{\frac{p}{1+p\|\boldsymbol{g}\|^2}}(\boldsymbol{u}_1^H\boldsymbol{h})\boldsymbol{u}_1 + \sqrt{p}\sum_{i=2}^{N_A}(\boldsymbol{u}_i^H\boldsymbol{h})\boldsymbol{u}_i\right)^H - p\frac{\boldsymbol{g}\boldsymbol{g}^H}{1+p\|\boldsymbol{g}\|^2}\end{aligned} \tag{14}$$

Instead, $\boldsymbol{\Lambda}^{1/2} = \text{diag}\left(\frac{1}{\sqrt{1+p\|\boldsymbol{g}\|^2}}, 1, \ldots, 1\right)$. Accordingly, the maximum eigenvalue of (13) is given by $1 + \lambda_Q(p)$, with $\lambda_Q(p)$ being the maximum eigenvalue of the matrix $\boldsymbol{Q}$ in (14), which can be computed as a function of $p$ leveraging Lemma 1, with

$$\begin{aligned}\|\boldsymbol{x}\|^2 &= \left\|\sqrt{\frac{p}{1+p\|\boldsymbol{g}\|^2}}(\boldsymbol{u}_1^H\boldsymbol{h})\boldsymbol{u}_1 + \sqrt{p}\sum_{i=2}^{N_A}(\boldsymbol{u}_i^H\boldsymbol{h})\boldsymbol{u}_i\right\|^2 \\ &= \frac{p|\boldsymbol{g}^H\boldsymbol{h}|^2}{\|\boldsymbol{g}\|^2(1+p\|\boldsymbol{g}\|^2)} + p\sum_{i=2}^{N_A}|\boldsymbol{u}_i^H\boldsymbol{h}|^2\ , \\ \|\boldsymbol{y}\|^2 &= \left\|\sqrt{\frac{p}{1+p\|\boldsymbol{g}\|^2}}\boldsymbol{g}\right\|^2 = \frac{p\|\boldsymbol{g}\|^2}{1+p\|\boldsymbol{g}\|^2}\ , \\ |\boldsymbol{y}^H\boldsymbol{x}|^2 &= \frac{p}{1+p\|\boldsymbol{g}\|^2}\times \\ &\times\left|\boldsymbol{g}^H\left(\sqrt{\frac{p(\boldsymbol{u}_1^H\boldsymbol{h})}{1+p\|\boldsymbol{g}\|^2}}\boldsymbol{u}_1 + p\sum_{i=2}^{N_A}(\boldsymbol{u}_i^H\boldsymbol{h})\boldsymbol{u}_i\right)\right|^2 \\ &= \frac{p^2}{(1+p\|\boldsymbol{g}\|^2)^2}|\boldsymbol{g}^H\boldsymbol{h}|^2\ .\end{aligned} \tag{15}$$

Elaborating, we obtain

$$\lambda_Q(p) = \frac{1}{2}\left[\alpha f(p) + \omega p + \sqrt{(\alpha f(p) + \omega p)^2 + 4g\omega p f(p)}\right]\ , \tag{16}$$

with $\omega = \sum_{i=2}^{N_A}|\boldsymbol{u}_i^H\boldsymbol{h}|^2$, $g = \|\boldsymbol{g}\|^2$, $c = |\boldsymbol{g}^H\boldsymbol{h}|^2$, $\alpha = (c/g - g)$, and $f(p) = \frac{p}{1+pg}$. At this point we are ready to show the following result.

**Proposition 1.** *The objective (8a) is a strictly PC function of $p$ and the solution of Problem (8) is*

$$p^* = \min(P_{max}, \bar{p})\ , \tag{17}$$

*with $\bar{p}$ the unique stationary point of (8a).*

*Proof:* See Appendix B. ■

## B. SKEE maximization with perfect CSI

When the secrecy key-rate is considered, the objective to maximize is (6). In this case, let us first observe that the secret-key rate can be expressed as

$$R_{s,key} = \log\left(1 + \frac{p\boldsymbol{w}^H\boldsymbol{h}\boldsymbol{h}^H\boldsymbol{w}}{1+p\boldsymbol{w}^H\boldsymbol{g}\boldsymbol{g}^H\boldsymbol{w}}\right)\ . \tag{18}$$

Then, the optimal beamforming vector $\boldsymbol{w}$ is once again determined as the solution of a generalized eigenvector problem. Specifically, the optimal $\boldsymbol{w}$ is the dominant generalized eigenvector of the matrix pencil $(p\boldsymbol{h}\boldsymbol{h}^H, \boldsymbol{I}_{N_A} + p\boldsymbol{g}\boldsymbol{g}^H)$, or equivalently, the dominant eigenvector of the rank-one matrix $p(\boldsymbol{I}_{N_A}+p\boldsymbol{g}\boldsymbol{g}^H)^{-1/2}\boldsymbol{h}\boldsymbol{h}^H(\boldsymbol{I}_{N_A}+p\boldsymbol{g}\boldsymbol{g}^H)^{-1/2}$. The corresponding maximum eigenvalue can be seen to be

$$\begin{aligned}\lambda_Q(p) &= p\boldsymbol{h}^H(\boldsymbol{I}_{N_A} + p\boldsymbol{g}\boldsymbol{g}^H)^{-1}\boldsymbol{h} \\ &= p\boldsymbol{h}^H\left[\boldsymbol{I}_{N_A} - \frac{p\boldsymbol{g}\boldsymbol{g}^H}{1+p\|\boldsymbol{g}\|^2}\right]\boldsymbol{h} = p\|\boldsymbol{h}\|^2 - \frac{p^2|\boldsymbol{h}^H\boldsymbol{g}|^2}{1+p\|\boldsymbol{g}\|^2}\ ,\end{aligned} \tag{19}$$

and (6) becomes

$$\text{SKEE} = \frac{\log(1+\lambda_Q(p))}{\mu p + P_c}\ . \tag{20}$$

By computing the second derivative of $\lambda_Q(p)$, it is immediately seen that $\lambda_Q(p)$ is a strictly concave function of $p$. Then,

(20) is a strictly PC function whose maximizer subject to a maximum power constraint is

$$p^* = \min(P_{max}, \bar{p}) \ , \tag{21}$$

with $\bar{p}$ the unique stationary point of (20).

### C. SEE maximization with statistical CSI

Assume now that Alice does not know the actual channel realization of the channel to the eavesdropper $\boldsymbol{g}$, but only its statistics. In particular, we model $\boldsymbol{g}$ as a complex circular Gaussian random vector with zero-mean and identity covariance matrix. Then, the resource allocation problem is formulated as the optimization program,

$$\max_{\boldsymbol{w},p} \frac{\mathbb{E}_{\boldsymbol{g}}\left[\log\left(\frac{1+p\boldsymbol{w}^H\boldsymbol{h}\boldsymbol{h}^H\boldsymbol{w}}{1+p\boldsymbol{w}^H\boldsymbol{g}\boldsymbol{g}^H\boldsymbol{w}}\right)\right]}{\mu p + P_c} \tag{22a}$$

$$\text{s.t. } p \in [0, P_{max}] \tag{22b}$$

$$\|\boldsymbol{w}\| = 1 \ . \tag{22c}$$

To begin with, we rewrite the numerator of (22a) as

$$\log\left(1+p\boldsymbol{w}^H\boldsymbol{h}\boldsymbol{h}^H\boldsymbol{w}\right) - \mathbb{E}_{\boldsymbol{g}}\left[\log\left(1+p\boldsymbol{w}^H\boldsymbol{g}\boldsymbol{g}^H\boldsymbol{w}\right)\right] \ , \tag{23}$$

from which we can observe that the optimal beamforming policy is to set $\boldsymbol{w} = \boldsymbol{h}/\|\boldsymbol{h}\|$, because the isotropic distribution of the channel vector $\boldsymbol{g}$ makes the direction $\boldsymbol{w}$ irrelevant for the second term in (23). As a result, Problem (22) becomes

$$\max_{p} \frac{\mathbb{E}_{\boldsymbol{g}}\left[\log\left(\frac{1+p\|\boldsymbol{h}\|^2}{1+p|\boldsymbol{h}^H\boldsymbol{g}|^2/\|\boldsymbol{h}\|^2}\right)\right]}{\mu p + P_c} \tag{24a}$$

$$\text{s.t. } p \in [0; P_{max}] \ . \tag{24b}$$

The statistical average in (24a) can be computed in closed form following [12], and (24a) can be expressed as

$$\frac{\log(1+p\|\boldsymbol{h}\|^2) - E_1\left(\frac{1}{p}\right)e^{1/p}}{\mu p + P_c} \ , \tag{25}$$

with $E_1$ being the exponential integral function, as defined in [25]. It follows that the secrecy capacity is non-negative only for

$$\|\boldsymbol{h}\|^2 \geq \frac{1}{p}\left[\exp\left(e^{1/p}E_1(1/p)\right) - 1\right] \ . \tag{26}$$

Using [25, 5.1.20], the right-hand side (RHS) of (26) can be bounded as

$$\frac{\sqrt{1+2p}-1}{p} \leq \frac{1}{p}\left[\exp\left(e^{1/p}E_1(1/p)\right) - 1\right] \leq 1 \ , \tag{27}$$

which shows that a necessary condition for (25) to be non-negative is $\|\boldsymbol{h}\|^2 \geq \frac{\sqrt{1+2p}-1}{p}$, whereas (25) is guaranteed to be positive provided $\|\boldsymbol{h}\|^2 \geq 1$.

Next, we turn our attention to solving Problem (24). We start by providing the following lemma.

**Lemma 2.** *Define the function $z(p) = 1 + 1/p - (1/p^2 + 2/p)e^{1/p}E_1(1/p)$. For all $p > 0$, it holds $0 \leq z(p) \leq 1$.*

*Proof:* We start by showing that $z(p) \geq 0$. Using [25, 5.1.14] and [25, 5.1.19], we have

$$E_1(1/p) = p(e^{-1/p} - E_2(1/p)) \ , \tag{28}$$

$$e^{1/p}E_1(1/p) = p(1 - e^{1/p}E_2(1/p)) \leq \frac{p^2+p}{1+2p} \ . \tag{29}$$

Then, it holds

$$z(p) \geq 1 + 1/p - (1/p^2 + 2/p)\frac{p^2+p}{1+2p} = 0 \ . \tag{30}$$

Finally, using again [25, 5.1.19] we obtain

$$\begin{aligned} 1 - z(p) &= -1/p + (1/p^2 + 2/p)e^{1/p}E_1(1/p) \\ &\geq -1/p + (1/p^2 + 2/p)\frac{p}{1+p} = \frac{1}{1+p} \geq 0 \ . \end{aligned} \tag{31}$$

■

Let us now define the function

$$y(p) = \frac{z(p) + \sqrt{z(p)}}{p(1-z(p))} \ , \tag{32}$$

which is well-defined and non-negative because of Lemma 2. The behavior of (32) is illustrated in Fig. 2, from which we see that it is a decreasing function for positive $p$ and that it tends to a finite value for $p \to 0$. Bearing this in mind, the next result provides a condition such that (25) is strictly PC.

**Proposition 2.** *For any given interval $[P_{min}, P_{max}]$, with $0 < P_{min} < P_{max}$, the function in (25) is a strictly PC function for all $p \in [P_{min}, P_{max}]$, if*

$$\|\boldsymbol{h}\|^2 > y(P_{min}) \ . \tag{33}$$

*In this case, the maximizer of (25) in the interval $[P_{min}, P_{max}]$ is given by*

$$p^* = \max(P_{min}, \min(P_{max}, \bar{p})) \ , \tag{34}$$

*with $\bar{p}$ the unique stationary point of (25).*

*Proof:* We recall the following formula from [25, p. 230]

$$\frac{de^xE_1(x)}{dx} = e^xE_1(x) - \frac{1}{x} \ . \tag{35}$$

Exploiting (35), we can compute the second derivative of the numerator of (25) as

$$\frac{z(p)}{p^2} - \frac{\|\boldsymbol{h}\|^4}{(1+p\|\boldsymbol{h}\|^2)^2} \ . \tag{36}$$

Elaborating, we obtain that (36) is negative if

$$(1-z(p))p^2\|\boldsymbol{h}\|^4 - 2\|\boldsymbol{h}\|^2pz(p) - z(p) > 0 \ . \tag{37}$$

Solving with respect to $\|\boldsymbol{h}\|^2$, we obtain the condition $\|\boldsymbol{h}\|^2 > y(p)$. Then, recalling that $y(p)$ is decreasing for positive $p$, we have that (25) is strictly concave for all $p \in [P_{min}, P_{max}]$, provided (33) holds. Finally, (24a) is the ratio between a strictly concave and an affine function, and is therefore strictly PC, thus admitting a unique stationary point which coincides with its global maximizer. ■

As a final remark, we observe how Fig. 2 shows that $y(p)$ can take larger values than 1 for small values of $p$. This means that, even if $\|\boldsymbol{h}\|^2 \geq 1$ was sufficient to ensure the non-negativity of (24a), it is not sufficient to guarantee that (24a) is also PC.

### D. SKEE maximization with statistical CSI

In this scenario, the problem to solve is stated as

$$\max_{\boldsymbol{w},p} \frac{\mathbb{E}_{\boldsymbol{g}}\left[\log\left(\frac{1+p\boldsymbol{w}^H\left(\boldsymbol{h}\boldsymbol{h}^H+\boldsymbol{g}\boldsymbol{g}^H\right)\boldsymbol{w}}{1+p\boldsymbol{w}^H\boldsymbol{g}\boldsymbol{g}^H\boldsymbol{w}}\right)\right]}{\mu p+P_c} \tag{38a}$$

$$\text{s.t. } p\in[0,P_{max}] \tag{38b}$$

$$\|\boldsymbol{w}\|=1\ . \tag{38c}$$

The optimal beamforming vector is once again $\boldsymbol{w}=\boldsymbol{h}/\|\boldsymbol{h}\|$, which results in the power control problem

$$\max_{p} \frac{\mathbb{E}_{\boldsymbol{g}}\left[\log\left(1+\frac{p\|\boldsymbol{h}\|^2}{1+p|\boldsymbol{h}^H\boldsymbol{g}|^2/\|\boldsymbol{h}\|^2}\right)\right]}{\mu p+P_c} \tag{39a}$$

$$\text{s.t. } p\in[0,P_{max}]\ . \tag{39b}$$

The function inside the expectation in (39a) can be checked to be strictly concave for any channel realization $\boldsymbol{g}$. Therefore, the numerator of (39a) is strictly concave as it is the expectation of strictly concave functions, which ultimately implies that (39a) is a strictly PC function. Thus, the solution to Problem (39) is expressed as

$$p^*=\min(P_{max},\bar{p})\ , \tag{40}$$

with $\bar{p}$ the unique stationary point of (39a).

## IV. SEE maximization in MIMO-ME systems

In this section we turn our attention to the general case in which an arbitrary number of antennas is deployed at all nodes. The structure of this section is similar to that for the MISO-SE case. In Sections IV-A and IV-B, perfect CSI is assumed, and the maximization of (1) and (2) is carried out, while Sections IV-C and IV-D deal with the corresponding maximizations with statistical CSI. In order to tackle the resulting optimization problems, we will extensively employ tools from fractional programming, which are introduced in Appendix A.

### A. SEE maximization with perfect CSI

In this scenario, the resource optimization problem to solve is formulated as

$$\max_{\boldsymbol{Q}\succeq 0} \frac{\log\left|\boldsymbol{I}_{N_B}+\boldsymbol{H}\boldsymbol{Q}\boldsymbol{H}^H\right|-\log\left|\boldsymbol{I}_{N_E}+\boldsymbol{G}\boldsymbol{Q}\boldsymbol{G}^H\right|}{\mu\text{tr}(\boldsymbol{Q})+P_c} \tag{41a}$$

$$\text{s.t. tr}(\boldsymbol{Q})\in[0,P_{max}] \tag{41b}$$

The challenge in solving (41) is that the numerator of (41a) is not concave in general, being the difference of concave functions. In this case, as argued in Appendix A, directly using fractional programming tools would result in an exponential complexity. For this reason, in the following we will trade-off complexity with optimality, developing two low-complexity algorithms, which also enjoy pleasant optimality claims.

The first algorithm will integrate fractional programming with the sequential convex optimization framework [26]–[30], and will be guaranteed to converge to a $\boldsymbol{Q}^*$ fulfilling the KKT optimality conditions of (41) by solving a sequence of fractional maximization problems with polynomial complexity.

The second algorithm, which will be referred to as *eigenmode selection* in the following, has an even lower computational complexity, requiring the solution of only one fractional problem with polynomial complexity. However, the resulting $\boldsymbol{Q}^*$ will enjoy a weaker optimality claim, being guaranteed to fulfill only a necessary condition such that the KKT conditions of (41) are satisfied.

*1) Sequential convex optimization:* The idea of sequential convex programming is to tackle a difficult maximization problem $\mathcal{P}$ with objective $f$, by solving a sequence of approximate problems $\{\mathcal{P}_\ell\}_\ell$ with objectives $\{f_\ell\}_\ell$, such that the following three properties are fulfilled, for all $\ell$:

**(P1)** $f_\ell(\boldsymbol{x})\le f(\boldsymbol{x})$, for all $\boldsymbol{x}$;
**(P2)** $f_\ell(\boldsymbol{x}^{(\ell-1)})=f(\boldsymbol{x})$, with $\boldsymbol{x}^{(\ell-1)}$ the maximizer of $f_{\ell-1}$;
**(P3)** $\nabla f_\ell(\boldsymbol{x}^{(\ell-1)})=\nabla f(\boldsymbol{x})$

Now, consider the sequences $\{\boldsymbol{x}^{(\ell)}\}_\ell$ of the solutions of the $\ell$-th Problem $\mathcal{P}_\ell$ of the sequence, and $\{f(\boldsymbol{x}_\ell)\}_\ell$ formed by the values of the objective of the original Problem $\mathcal{P}$, evaluated at $\boldsymbol{x}_\ell$. If Properties (**P1**), (**P2**), and (**P3**) are fulfilled, both sequences $\{\boldsymbol{x}^{(\ell)}\}_\ell$ and $\{f(\boldsymbol{x}_\ell)\}_\ell$ converge. Moreover, $\{f(\boldsymbol{x}_\ell)\}_\ell$ is monotonically increasing, while $\{\boldsymbol{x}^{(\ell)}\}_\ell$ converges to a point fulfilling the KKT optimality conditions of the original Problem $\mathcal{P}$ [26].

Therefore, we can determine a matrix $\boldsymbol{Q}$ which fulfills the KKT equations of (41), provided we can find suitable lower-bounds of (41a) which also fulfill Properties (**P2**) and (**P3**). Moreover, the lower-bounds should be easier to maximize than (41a). To this end, we proceed as follows. Exploiting the fact that the numerator of (41a) is the difference of two concave functions, for any given $\boldsymbol{Q}_0$ we can obtain a lower-bound of (41) by replacing the second summand in the numerator with its first-order Taylor expansion[3]. Then, for any $\boldsymbol{Q}_0\succeq\boldsymbol{0}$ we have

$$\begin{aligned}\text{SEE}&\ge\widetilde{\text{SEE}}\\&=\frac{\log\left|\boldsymbol{I}_{N_B}+\boldsymbol{H}\boldsymbol{Q}\boldsymbol{H}^H\right|-\left(g_0+2\Re\left\{\text{tr}\left(\boldsymbol{M}_0^H(\boldsymbol{Q}-\boldsymbol{Q}_0)\right)\right\}\right)}{\mu\text{tr}(\boldsymbol{Q})+P_c}\end{aligned} \tag{42}$$

with $g_0=\log\left|\boldsymbol{I}_{N_E}+\boldsymbol{G}\boldsymbol{Q}_0\boldsymbol{G}^H\right|$, $\boldsymbol{M}_0=\boldsymbol{G}^H\left(\boldsymbol{I}_{N_E}+\boldsymbol{G}\boldsymbol{Q}_0\boldsymbol{G}^H\right)^{-1}\boldsymbol{G}$ and where we have exploited the fact that the conjugate gradient of $\log\left|\boldsymbol{I}_{N_E}+\boldsymbol{G}\boldsymbol{Q}\boldsymbol{G}^H\right|$ is the Hermitian matrix $\boldsymbol{G}^H\left(\boldsymbol{I}_{N_E}+\boldsymbol{G}\boldsymbol{Q}_0\boldsymbol{G}^H\right)^{-1}\boldsymbol{G}$ [31]. Clearly, $\text{SEE}(\boldsymbol{Q}_0)=\widetilde{\text{SEE}}(\boldsymbol{Q}_0)$ and $\nabla\text{SEE}(\boldsymbol{Q}_0)=\nabla\widetilde{\text{SEE}}(\boldsymbol{Q}_0)$, for any $\boldsymbol{Q}_0\succeq\boldsymbol{0}$, so that all three Properties (**P1**), (**P2**), and (**P3**) are fulfilled. Thus, in the sequential convex optimization framework, the $\ell$-th problem $\mathcal{P}_\ell$ of the sequence

[3] We recall that given a real-valued function $g$ of complex, matrix argument $\boldsymbol{Q}$, the first-order Taylor expansion around a point $\boldsymbol{Q}_0$ is written as [31], [32] $g(\boldsymbol{Q})=g(\boldsymbol{Q}_0)+2\Re\left\{\text{tr}\left(\left(\nabla_{\boldsymbol{Q}^*}g\right)^H\Big|_{\boldsymbol{Q}=\boldsymbol{Q}_0}(\boldsymbol{Q}-\boldsymbol{Q}_0)\right)\right\}$.

takes the form

$$(\mathcal{P}_\ell): \quad \max_{\boldsymbol{Q}\succeq 0} \ \widetilde{\text{SEE}} \qquad (43a)$$

$$\text{s.t. } \text{tr}(\boldsymbol{Q}) \in [0, P_{max}] \ . \qquad (43b)$$

The objective (43a) has a concave numerator and an affine denominator and so it can be globally optimized by fractional programming with polynomial complexity. Then, sequential convex optimization and fractional programming can be integrated to determine a KKT point of (41). The overall algorithm is illustrated in Algorithm 1, and works by updating in each iteration the point $\boldsymbol{Q}_0$ around which the Taylor expansion is computed, and then solving $(\mathcal{P}_\ell)$ using fractional programming. We remark that in principle any fractional programming method can be used to globally solve $(\mathcal{P}_\ell)$ with polynomial complexity. In Algorithm 1, and also in the rest of this paper, without loss of generality, we resort to Dinkelbach's algorithm, which is the most commonly used fractional programming algorithm. More details on Dinkelbach's algorithm are reported in Appendix A.

**Algorithm 1** Algorithm for SEE maximization in MIMO-ME systems with perfect CSI.

$\ell = 0$; $\epsilon > 0$; Select a feasible $\boldsymbol{Q}_0^{(\ell)}$;
**while** $\left|\text{SEE}\left(\boldsymbol{Q}_0^{(\ell)}\right) - \text{SEE}\left(\boldsymbol{Q}_0^{(\ell-1)}\right)\right| > \epsilon$ **do**
  Compute $\boldsymbol{M}_0$ and $g_0$;
  Solve Problem (43) by Dinkelbach's algorithm and set $\boldsymbol{Q}$ as the solution.
  $\boldsymbol{Q}_0^{(\ell)} = \boldsymbol{Q}$;
  $\ell = \ell + 1$;
**end while**

Following a similar reasoning as in [26], we can prove the following result.

**Proposition 3.** *After each iteration of Algorithm 1, the value of the true SEE is not decreased. Moreover, Algorithm 1 is guaranteed to converge to a* $\boldsymbol{Q}^*$ *which fulfills the KKT optimality conditions of Problem (41).*

**Remark 1.** *We observe that Algorithm 1 can be readily specialized to perform secrecy rate maximization, by simply setting* $\mu = 0$ *and* $P_c = 1$. *In this case Problem (43) becomes a non-fractional, concave, problem, which can be solved by standard methods, and a similar result as Proposition 3 holds.*

*2) Eigenmode selection:* This approach is inspired to a result first presented in [6]. There, the problem of secrecy capacity maximization in MIMO-ME systems is studied and a transmission scheme is proposed such that the resulting $\boldsymbol{Q}$ fulfills a necessary condition for the KKT conditions of the secrecy capacity maximization problem to hold [6, Theorem 1]. In the following we will extend this result to the problem of SEE maximization. The first step is represented by the following result, which extends [6, Theorem 1] to the case of SEE.

**Proposition 4.** *Define the EVD of* $\boldsymbol{Q}$ *as* $\boldsymbol{U}_Q\boldsymbol{\Lambda}_Q\boldsymbol{U}_Q^H$, *and let* $\boldsymbol{u}_{+,Q}$ *be an eigenvector of* $\boldsymbol{Q}$ *corresponding to a positive eigenvalue. Then, in order to fulfill the KKT conditions of (41)* $\boldsymbol{u}_{+,Q}$ *must satisfy*

$$\boldsymbol{u}_{+,Q}^H(\boldsymbol{H}^H\boldsymbol{H} - \boldsymbol{G}^H\boldsymbol{G})\boldsymbol{u}_{+,Q} > 0 \ . \qquad (44)$$

*Proof:* See Appendix C. ■

The next step is to observe that a way to obtain a covariance matrix $\boldsymbol{Q}$ that fulfills (44) is to solve the following optimization problem.

$$\max_{\boldsymbol{Q}\succeq 0} \ \frac{\log|\boldsymbol{I}_{N_B} + \boldsymbol{H}_+\boldsymbol{Q}| - \log|\boldsymbol{I}_{N_E} + \boldsymbol{G}_+\boldsymbol{Q}|}{\mu\text{tr}(\boldsymbol{Q}) + P_c} \ , \qquad (45a)$$

$$\text{s.t. } \text{tr}(\boldsymbol{Q}) \leq P_{max} \ , \qquad (45b)$$

with $\boldsymbol{H}_+$ and $\boldsymbol{G}_+$ the projection of $\boldsymbol{H}^H\boldsymbol{H}$ and $\boldsymbol{G}^H\boldsymbol{G}$ onto the positive eigenspace of $\boldsymbol{H}^H\boldsymbol{H} - \boldsymbol{G}^H\boldsymbol{G}$, i.e. $\boldsymbol{H}_+ = \boldsymbol{P}_+\boldsymbol{H}^H\boldsymbol{H}\boldsymbol{P}_+$ and $\boldsymbol{G}_+ = \boldsymbol{P}_+\boldsymbol{G}^H\boldsymbol{G}\boldsymbol{P}_+$, with $\boldsymbol{P}_+ = \boldsymbol{U}_+\boldsymbol{U}_+^H$, and $\boldsymbol{U}_+$ the semi-unitary matrix whose columns are the eigenvectors of $\boldsymbol{H}^H\boldsymbol{H} - \boldsymbol{G}^H\boldsymbol{G}$ corresponding to positive eigenvalues. To see why this is true, we can observe that (45a) can be equivalently rewritten as

$$\frac{\log\left|\boldsymbol{I}_{N_B} + \boldsymbol{H}^H\boldsymbol{H}\boldsymbol{P}_+\boldsymbol{Q}\boldsymbol{P}_+\right| - \log\left|\boldsymbol{I}_{N_E} + \boldsymbol{G}^H\boldsymbol{G}\boldsymbol{P}_+\boldsymbol{Q}\boldsymbol{P}_+\right|}{\mu\text{tr}(\boldsymbol{Q}) + P_c} \ , \qquad (46)$$

from which it follows that the solution of (45a) must have positive eigenvectors spanning the space spanned by $\boldsymbol{P}_+$. Indeed, if the optimal $\boldsymbol{Q}_+^*$ had some eigenvectors outside the space spanned by $\boldsymbol{P}_+$, then $\boldsymbol{Q}_+^* \neq \boldsymbol{P}_+\boldsymbol{Q}_+^*\boldsymbol{P}_+$. However, plugging $\boldsymbol{Q}_+^*$ or $\boldsymbol{P}_+\boldsymbol{Q}_+^*\boldsymbol{P}_+$ in the numerator of (46) yields the same value, whereas $\text{tr}(\boldsymbol{Q}_+^*) > \text{tr}(\boldsymbol{P}_+\boldsymbol{Q}_+^*\boldsymbol{P}_+)$. Therefore $\boldsymbol{Q}_+^*$ would yield a larger denominator, and thus a smaller objective, thus contradicting the assumption that $\boldsymbol{Q}_+^*$ is the global optimum. So, the positive eigenvectors of $\boldsymbol{Q}_+^*$ belong to the space spanned by $\boldsymbol{P}_+$, which coincides with the space spanned by $\boldsymbol{U}_+$. Therefore, the positive eigenvectors of $\boldsymbol{Q}_+^*$ fulfill (44).

Focusing now on the transformed Problem (45), we observe that the transformed channels $\boldsymbol{H}_+$ and $\boldsymbol{G}_+$ enjoy by construction the property that $\boldsymbol{H}_+ \succeq \boldsymbol{G}_+$. Under this condition, the numerator of (45a) is known to be a concave function [6], and therefore (45a) can be globally maximized by means of fractional programming tools, such as Dinkelbach's algorithm. The solution $\boldsymbol{Q}_+^*$ will satisfy (44), which is a necessary condition to fulfill the KKT conditions of the original problem (41). This shows the trade-off between the eigen-mode selection and Algorithm 1. The first method is computationally simpler, but, unlike Algorithm 1, can not ensure to obtain a KKT point of (41).

**Remark 2.** *From the discussion above, it follows that if the original channels* $\boldsymbol{H}$ *and* $\boldsymbol{G}$ *are such that* $\boldsymbol{H}\boldsymbol{H}^H \succeq \boldsymbol{G}\boldsymbol{G}^H$, *then the eigenmode selection scheme achieves the global optimum of the SEE. In this scenario, Algorithm 1 is also guaranteed to achieve global optimality, because if* $\boldsymbol{H}\boldsymbol{H}^H \succeq \boldsymbol{G}\boldsymbol{G}^H$, *then Problem (7) has a concave numerator, and hence a PC objective, thus implying that KKT conditions are necessary and sufficient optimality conditions.*

### B. SKEE maximization with perfect CSI

In this case, we have to consider the following optimization problem

$$\max_{\boldsymbol{Q}\succeq 0} \frac{\log\left|\boldsymbol{I}_{N_A}+\boldsymbol{Q}(\boldsymbol{H}^H\boldsymbol{H}+\boldsymbol{G}^H\boldsymbol{G})\right|-\log\left|\boldsymbol{I}_{N_A}+\boldsymbol{Q}\boldsymbol{G}^H\boldsymbol{G}\right|}{\mu \mathrm{tr}(\boldsymbol{Q})+P_c} \tag{47a}$$

$$\text{s.t. } \mathrm{tr}(\boldsymbol{Q})\in[0,P_{max}]\ . \tag{47b}$$

Since it holds that $\boldsymbol{H}^H\boldsymbol{H}+\boldsymbol{G}\boldsymbol{G}^H \succeq \boldsymbol{G}\boldsymbol{G}^H$ we have that, as observed in Section IV-A, the numerator of (47a) is a concave function of $\boldsymbol{Q}$. This in turn guarantees that (47a) is a PC of $\boldsymbol{Q}$, given that the denominator is clearly affine in $\boldsymbol{Q}$. Thus, Problem (47) can be globally solved with polynomial complexity by means of fractional programming tools, such as Dinkelbach's algorithm.

### C. SEE maximization with statistical CSI

In this case, the resource allocation problem is formulated as

$$\max_{\boldsymbol{Q}\succeq 0} \frac{\log\left|\boldsymbol{I}_{N_B}+\boldsymbol{H}\boldsymbol{Q}\boldsymbol{H}^H\right|-\mathbb{E}_{\boldsymbol{G}}\left[\log\left|\boldsymbol{I}_{N_E}+\boldsymbol{G}\boldsymbol{Q}\boldsymbol{G}^H\right|\right]}{\mu \mathrm{tr}(\boldsymbol{Q})+P_c}\ , \tag{48a}$$

$$\text{s.t. } \mathrm{tr}(\boldsymbol{Q})\in[0,P_{max}]\ , \tag{48b}$$

with $\boldsymbol{G}$ a random matrix with independent and identically distributed (i.i.d.), Gaussian, zero-mean entries, with unit-variance.

Due to its isotropic distribution, multiplying $\boldsymbol{G}$, by left or right, by a unitary matrix does not change its distribution. As a consequence, we have that the eigenvectors of $\boldsymbol{Q}$ do not affect the second summand at the numerator of (48a). Moreover, they do not affect the denominator, too. Then, the eigenvectors of $\boldsymbol{Q}$ are to be chosen so as to maximize the first summand at the numerator, which means that $\boldsymbol{Q}$ should diagonalize the legitimate channel. Upon doing this, (48) becomes

$$\max_{\{q_i\geq 0\}_i} \frac{\sum_{i=1}^{N_A}\log(1+q_ih_i)-\mathbb{E}\left[\log\left|\boldsymbol{I}_{N_E}+\sum_{i=1}^{N_A}q_i\boldsymbol{g}_i\boldsymbol{g}_i^H\right|\right]}{\mu\sum_{i=1}^{N_A}q_i+P_c} \tag{49a}$$

$$\text{s.t } \sum_{i=1}^{N_A}q_i\leq P_{max} \tag{49b}$$

$$q_i\geq 0\ ,\forall\ ,i=1,\ldots,N_A\ , \tag{49c}$$

wherein, for all $i=1,\ldots,N_A$, $q_i$ is the $i$-th eigenvalue of $\boldsymbol{Q}$, $h_i$ is the $i$-th largest eigenvalue of $\boldsymbol{H}\boldsymbol{H}^H$, and the statistical expectation is with respect to $\boldsymbol{g}_i$, for all $i=1,\ldots,N_E$, with $\boldsymbol{g}_i$ the $i$-th column of $\boldsymbol{G}$. Problem (49) is still difficult to solve because, as in the perfect CSI case, the numerator of the objective is in general not concave. However, we can apply a similar approach as in the perfect CSI scenario, merging fractional programming with sequential convex optimization to determine a KKT point of (49). Again, we will accomplish this by computing the first-order Taylor expansion of the second summand at the numerator of (49a). To this end, let us define

$$g(\boldsymbol{q})=\mathbb{E}_{\{\boldsymbol{g}_i\}_i}\left[\log\left|\boldsymbol{I}_{N_E}+\sum_{i=1}^{N_A}q_i\boldsymbol{g}_i\boldsymbol{g}_i^H\right|\right]\ , \tag{50}$$

and, for all $i=1,\ldots,N_A$, $\boldsymbol{Z}_i=\sum_{j\neq i}q_j\boldsymbol{g}_j\boldsymbol{g}_j^H+\boldsymbol{I}_{N_E}$. Next, the partial derivative of $g$ with respect to the generic $q_k$ can be computed exploiting the following formula [33],

$$\frac{d\log|\boldsymbol{A}+x\boldsymbol{B}|}{dx}=\mathrm{tr}\left((\boldsymbol{A}+x\boldsymbol{B})^{-1}\boldsymbol{B}\right)\ . \tag{51}$$

Then, we have

$$\begin{aligned}\frac{\partial g}{\partial q_k}&=\mathbb{E}_{\{\boldsymbol{g}_i\}_i}\left[\boldsymbol{g}_k^H(\boldsymbol{Z}_k+q_k\boldsymbol{g}_k\boldsymbol{g}_k^H)^{-1}\boldsymbol{g}_k\right]\\&=\mathbb{E}_{\{\boldsymbol{g}_i\}_i}\left[\boldsymbol{g}_k^H\left(\boldsymbol{Z}_k^{-1}-\frac{q_k\boldsymbol{Z}_k^{-1}\boldsymbol{g}_k\boldsymbol{g}_k^H\boldsymbol{Z}_k^{-1}}{1+q_k\boldsymbol{g}_k^H\boldsymbol{Z}_k^{-1}\boldsymbol{g}_k}\right)\boldsymbol{g}_k\right]\\&=\frac{1}{q_k}\left(1-\mathbb{E}_{c_k}\left[\frac{1}{1+q_kc_k}\right]\right)\ ,\end{aligned} \tag{52}$$

wherein $c_k=\boldsymbol{g}_k^H\boldsymbol{Z}_k^{-1}\boldsymbol{g}_k$. Then, for any feasible $\boldsymbol{q}_0=\{q_{0,i}\}_{i=1}^{N_A}$, the approximate problem to be considered in each iteration of the sequential convex optimization algorithm is expressed as

$$\max_{\{q_i\geq 0\}_i} \frac{\sum_{i=1}^{N_A}\log(1+q_ih_i)-g(\boldsymbol{q}_0)-\sum_{i=1}^{N_A}(q_i-q_{0,i})\frac{\partial g}{\partial q_i}\Big|_{\boldsymbol{q}=\boldsymbol{q}_0}}{\mu\sum_{i=1}^{N_A}q_i+P_c} \tag{53a}$$

$$\text{s.t. } \sum_{i=1}^{N_A}q_i\leq P_{max} \tag{53b}$$

$$q_i\geq 0\ ,\forall\ i=1,\ldots,N_A\ . \tag{53c}$$

The numerator and denominator of (53a) are respectively concave and affine in $\boldsymbol{q}$ and therefore (53) can be solved by fractional programming with polynomial complexity. The complete resource allocation algorithm can be formulated as in the following Algorithm 2, which enjoys similar properties as Algorithm 1.

**Algorithm 2** Algorithm for SEE maximization in MIMO-ME systems with statistical CSI.

$\ell=0$; $\epsilon>0$; Select a feasible $\boldsymbol{q}_0^{(\ell)}$;
**while** $\left|\mathrm{SEE}\left(\boldsymbol{q}_0^{(\ell)}\right)-\mathrm{SEE}\left(\boldsymbol{q}_0^{(\ell-1)}\right)\right|>\epsilon$ **do**
  Compute $\frac{\partial g}{\partial q_i}\Big|_{\boldsymbol{q}=\boldsymbol{q}_0}$, for all $i=1,\ldots,N_A$, as in (52);
  Solve (53) by Dinkelbach's algorithm, and set $\boldsymbol{q}$ as the solution;
  $\boldsymbol{q}_0^{(\ell)}=\boldsymbol{q}$;
  $\ell=\ell+1$;
**end while**

### D. SKEE maximization with statistical CSI

In this scenario, the resource allocation problem to address is formulated as

$$\max_{\boldsymbol{Q}\succeq 0} \frac{\mathbb{E}_{\boldsymbol{G}}\left[\log\left(\frac{\left|\boldsymbol{I}_{N_A}+\left(\boldsymbol{H}^H\boldsymbol{H}+\boldsymbol{G}^H\boldsymbol{G}\right)\boldsymbol{Q}\right|}{\left|\boldsymbol{I}_{N_A}+\boldsymbol{G}^H\boldsymbol{G}\boldsymbol{Q}\right|}\right)\right]}{\mu\text{tr}(\boldsymbol{Q})+P_c}, \tag{54a}$$

$$\text{s.t. } \text{tr}(\boldsymbol{Q}) \in [0, P_{max}] \,. \tag{54b}$$

Recalling the discussion in Section IV-B, we observe that the argument of the statistical average in (54a) is a concave function for any realization of the channel matrix $\boldsymbol{G}$, since we always have $\boldsymbol{H}^H\boldsymbol{H}+\boldsymbol{G}^H\boldsymbol{G} \succeq \boldsymbol{G}^H\boldsymbol{G}$. This in turn implies that the numerator of (54a) is a concave function, because of the linearity of the expectation operator. Thus, (54) is a PC maximization problem which can be globally and efficiently maximized by directly applying fractional programming tools.

## V. Numerical Results

We focus first on the MISO-SE scenario. Figs. 3 and 4 respectively illustrate the instantaneous SEE (5) and secrecy rate versus $P_{max}$, achieved by the following resource allocation schemes:

- SEE maximization with perfect CSI, as described in Section III-A;
- SEE maximization with statistical CSI, as described in Section III-C;
- secrecy rate maximization with perfect CSI;
- maximum power allocation, i.e. $p = P_{max}$, and beamformer $\boldsymbol{w} = \boldsymbol{h}/\|\boldsymbol{h}\|$ matched to the legitimate channel.

For any channel realization $(\boldsymbol{h}, \boldsymbol{g})$, the transmit power $p$ and beamforming vector $\boldsymbol{w}$ resulting from each algorithm have been used to compute (5), and the presented results have been obtained by averaging over 1000 independent channel realizations, generated as Gaussian random vectors with zero mean and identity covariance matrix. The system parameters have been set to $P_c = 5$W, $\mu = 1$, and $N_A = 3$, while the communication bandwidth is 1MHz.

As expected, from Fig. 3 we find that the best SEE is obtained when perfect CSI is available, whereas a gap is observed in case only statistical CSI is used for resource allocation. Moreover, we can observe how the SEE achieved by SEE optimization eventually saturates for high $P_{max}$. This is explained recalling that the SEE is not monotonically increasing with the transmit power, but instead admits a maximum value. Therefore, for low values of $P_{max}$, transmitting with full power is the optimal energy-efficient policy, but once $P_{max}$ is large enough to allow achieving the peak value of the SEE, the excess transmit power is not used and the achieved SEE value keeps constantly equal to the peak value. Instead, for large $P_{max}$, the SEE achieved by the resource allocation that maximizes the secrecy rate decreases, because in this case all of the available power $P_{max}$ is used.

A similar scenario is considered in Fig. 4, with the difference that the reported metric is the secrecy rate rather than the SEE. In this case, as expected, allocating the resource so as to maximize the secrecy rate yields an increasing achieved value of the secrecy rate. Instead, the value of the secrecy rate achieved by maximizing the SEE eventually saturates, both with perfect and statistical CSI, which can be explained by similar remarks as for Fig. 3.

Next, we analyze the MIMO-ME scenario. The considered system parameters are $N_A = N_B = N_E = 2$, $P_c = 5$W, and $\mu = 1$. Each entry of the channel matrices has been generated as a realization of a zero-mean Gaussian random variable with variance $\sigma_h^2$ and $\sigma_g^2$ for the legitimate and eavesdropper channels, respectively. The following illustrations have been obtained by averaging over 1000 independent channel realizations.

Fig. 5 compares the achieved instantaneous SEE versus $P_{max}$, with $\sigma_h = 2$ and $\sigma_g = 1$, for the following resource allocation schemes:

- SEE maximization with perfect CSI by Algorithm 1;
- SEE maximization with perfect CSI and eigen-mode selection from [6];
- Secrecy rate maximization with perfect CSI by adapted Algorithm 1 as explained in Remark 1;
- SEE maximization with statistical CSI by Algorithm 2.

As in the MISO-SE case, the best performance is obtained when perfect CSI is available. Moreover, the results indicate that the proposed Algorithm 1, outperforms the eigenmode selection scheme which allocates the transmit directions after [6]. This shows the trade-off between Algorithm 1, which is able to determine a KKT point of the SEE maximization problem by solving a sequence of PC fractional problems, and the more heuristic eigenmode selection scheme, which requires to solve only one PC fractional problem, but can not claim to obtain a KKT point of the problem.

The comparison between Algorithm 1 and the eigenmode selection scheme is further analyzed in Fig. 6. Here, we compare the SEE achieved by the two schemes for increasing values of $\sigma_h = 2; 6; 10$ and $\sigma_g = 1$. We can see that the gap between the two schemes shrinks for increasing $\sigma_h$. This is explained observing that increasing $\sigma_h$ scales the eigenvalues of $\boldsymbol{H}\boldsymbol{H}^H$ by a factor $\sigma_h^2$, and that $\boldsymbol{H}\boldsymbol{H}^H$ is full-rank with probability[4] 1. As a consequence, for increasing $\sigma_h$ the probability that $\boldsymbol{H}\boldsymbol{H}^H \succeq \boldsymbol{G}\boldsymbol{G}^H$ increases, and, as observed in Remark 2, in this case both algorithms globally maximize the SEE.

Finally, we analyze the performance of the proposed sequential convex optimization algorithm in terms of obtained secrecy rate. In particular, Fig. 7 illustrates the secrecy rate obtained by Algorithm 1 when it is used to optimize both the SEE and the secrecy rate, for $\sigma_h = 2$. The performance of Algorithm 1 in terms of secrecy rate are contrasted to the global optimum, which is provided in [8] for the special case of $N_A = 2$. Remarkably, the results indicate that the proposed sequential convex optimization framework virtually performs as the global optimum.

[4]Recall that both $\boldsymbol{H}\boldsymbol{H}^H$ and $\boldsymbol{G}\boldsymbol{G}^H$ are realizations of Wishart matrices, and that the probability that the minimum eigenvalue of a Wishart matrix with dimension $n$ is lower than some positive $z$ is equal to $1 - e^{-nz}$ [34].

## VI. Conclusion

In this work we have investigated the problem of resource allocation in MISO-SE and MIMO-ME systems. Unlike most previous related papers, the goal of the resource allocation process has been on maximizing the system bit/Joule energy efficiency. In particular, two energy-efficient metrics have been considered, the SEE, defined as the ratio between the secrecy rate and the consumed power, and the SKEE, defined as the ratio between the secret-key rate and the consumed power. The resource allocation has been performed considering both perfect CSI and statistical CSI of the channel between the legitimate user and the eavesdropper.

In the MISO-SE scenario, we show how the problems can be reformulated as PC maximizations, determining the globally optimal resource allocation.

In the MIMO-ME scenario, we use fractional programming to obtain the global optimum of the SKEE maximization problem. The SEE maximization problem is more challenging and we integrate fractional programming with sequential convex optimization to find resource allocations fulfilling first-order optimality conditions. Numerical results indicate that the proposed approach achieves near-optimal performance.

## Appendix A
## Generalized concavity and fractional programming theory

We limit our review to those results which are used in this work. For a more detailed review, we refer the reader to [35, Chapter 6], [36, Chapters 3, 4], [37], and [14].

**Definition 1** (Pseudo-concavity)**.** *Let $\mathcal{C} \subseteq \mathbb{R}^n$ be a convex set. Then $r : \mathcal{C} \to \mathbb{R}$ is PC if and only if, for all $\boldsymbol{x}_1, \boldsymbol{x}_2 \in \mathcal{C}$, it is differentiable and*

$$r(\boldsymbol{x}_2) < r(\boldsymbol{x}_1) \Rightarrow \nabla(r(\boldsymbol{x}_2))^T(\boldsymbol{x}_1 - \boldsymbol{x}_2) > 0 \ . \qquad (55)$$

In a similar way we can define strict pseudo-concavity.

**Definition 2** (Strict pseudo-concavity)**.** *Let $\mathcal{C} \subseteq \mathbb{R}^n$ be a convex set. Then $r : \mathcal{C} \to \mathbb{R}$ is strictly PC if and only if, for all $\boldsymbol{x}_1 \neq \boldsymbol{x}_2 \in \mathcal{C}$, it is differentiable and*

$$r(\boldsymbol{x}_2) \leq r(\boldsymbol{x}_1) \Rightarrow \nabla(r(\boldsymbol{x}_2))^T(\boldsymbol{x}_1 - \boldsymbol{x}_2) > 0 \ . \qquad (56)$$

The interest for PC functions stems from the following result.

**Proposition 5.** *Let $r : \mathcal{C} \to \mathbb{R}$ be a PC function. Then,*

(a) *If $\boldsymbol{x}^*$ is a stationary point for $r$, then it is a global maximizer for $r$;*
(b) *The KKT conditions for the problem of maximizing $r$ subject to convex constraints are necessary and sufficient conditions for optimality;*
(c) *If $r$ is strictly PC, then a unique maximizer exists.*

Pseudo-concavity plays a key-role in the optimization of fractional functions, due to the following result.

**Proposition 6.** *Let $r(\boldsymbol{x}) = \dfrac{f(\boldsymbol{x})}{g(\boldsymbol{x})}$, with $f : \mathcal{C} \subseteq \mathbb{R}^n \to \mathbb{R}$ and $g : \mathcal{C} \subseteq \mathbb{R}^n \to \mathbb{R}_+$. If $f$ is non-negative, differentiable, and concave, while $g$ is differentiable and convex, then $r$ is PC. If $g$ is affine, the non-negativity of $f$ can be relaxed. Strict pseudo-concavity holds if either $f$ is strictly concave, or $g$ is strictly convex.*

Finally, let us introduce the definition of fractional program.

**Definition 3** (Fractional program)**.** *Let $\mathcal{X} \subseteq \mathbb{R}^n$, and consider the functions $f : \mathcal{X} \to \mathbb{R}_0^+$ and $g : \mathcal{X} \to \mathbb{R}^+$. A fractional program is the optimization problem*

$$\max_{\boldsymbol{x} \in \mathcal{X}} \frac{f(\boldsymbol{x})}{g(\boldsymbol{x})} \ . \qquad (57)$$

The following result relates the solution of (57) to the auxiliary function $F(\beta) = \max_{\boldsymbol{x} \in \mathcal{X}} \{f(\boldsymbol{x}) - \beta g(\boldsymbol{x})\}$.

**Proposition 7** ( [37])**.** *An $\boldsymbol{x}^* \in \mathcal{X}$ solves* (57) *if and only if $\boldsymbol{x}^* = \arg\max_{\boldsymbol{x} \in \mathcal{X}} \{f(\boldsymbol{x}) - \beta^* g(\boldsymbol{x})\}$, with $\beta^*$ being the unique zero of $F(\beta)$. Moreover, $\beta^*$ coincides with the global maximum of* (57)*.*

This result allows us to solve (57) by finding the zero of $F(\beta)$. An efficient algorithm to do so is the Dinkelbach's algorithm, which exhibits a super-linear convergence rate [37], and is reported here in Algorithm 3.

**Algorithm 3** Dinkelbach's algorithm

Set $\varepsilon > 0$; $\beta = 0$; $F > \varepsilon$;
**while** $F \geq \varepsilon$ **do**
  $\boldsymbol{x}^* = \arg\max_{\boldsymbol{x} \in \mathcal{X}} \{f(\boldsymbol{x}) - \beta g(\boldsymbol{x})\}$
  $F = f(\boldsymbol{x}^*) - \beta g(\boldsymbol{x}^*)$;
  $\beta = f(\boldsymbol{x}^*)/\, g(\boldsymbol{x}^*)$;
**end while**

If $f(\boldsymbol{x})$ and $g(\boldsymbol{x})$ are concave and convex, respectively, and if $\mathcal{X}$ is a convex set, then the Dinkelbach's algorithm requires to solve one convex problem in each iteration. If instead one of these assumptions is not fulfilled, Dinkelbach's algorithm is still guaranteed to converge to the global solution, but the auxiliary problem to be *globally* solved in each iteration to compute $F$ becomes NP-hard. More in general, in this scenario no computationally-efficient algorithm is known to find the global solution of (57).

## Appendix B
## Proof of Proposition 1

The result follows if we can show that the objective function is the ratio of a strictly concave over an affine function. The denominator of the objective is clearly affine, and we are left to show the concavity of the numerator. To this end, we need to prove that $\lambda_Q(p)$ is concave in $p$. From (16) we can compute $\lambda_Q(p)$ as shown in (58) at the top of next page. Next, upon defining the function

$$z(p) = \sqrt{(\alpha + 2g + \omega(1 + pg))^2 - 4g(g + \alpha)} \qquad (59)$$

we have $\lambda_Q(p) = (f(p)(z(p) + \alpha) + \omega p)/2$, and the thesis follows if we can show that $f(p)(z(p) + \alpha)$ is concave in $p$. To this end, we can equivalently show that $gf(x)(z(x) + \alpha)$

$$\begin{aligned}\lambda_Q(p) &= \frac{1}{2}\left[\alpha f(p) + \omega p + \sqrt{(\alpha f(p) + \omega p)^2 + 4g\omega p f(p)}\right] \\ &= \frac{1}{2}\left[\alpha f(p) + \omega p + f(p)\sqrt{\alpha^2 + \omega^2(1+pg)^2 + 2\omega(\alpha + 2g)(1+pg)}\right] \\ &= \frac{1}{2}\left[\alpha f(p) + \omega p + f(p)\sqrt{(\alpha + 2g + \omega(1+pg))^2 - 4g(g+\alpha)}\right]\end{aligned} \tag{58}$$

---

$$\begin{aligned} z^{'}(x) &= \omega \frac{\omega(1+x) + \alpha + 2g}{\sqrt{(\omega(1+x) + \alpha + 2g)^2 - 4g(g+\alpha)}} \\ z^{''}(x) &= \frac{-4g(g+\alpha)\omega^2}{\sqrt{(\omega(1+x) + \alpha + 2g)^2 - 4g(g+\alpha)}\left((\omega(1+x) + \alpha + 2g)^2 - 4g(g+\alpha)\right)}\end{aligned} \tag{61}$$

---

is concave in $x = pg$. Computing the second derivative of $gf(x)(z(x) + \alpha)$, after ordinary elaborations we obtain

$$\frac{2z^{'}(x)}{(1+x)^2} - \frac{2(z(x)+\alpha)}{(1+x)^3} + \frac{xz^{''}(x)}{1+x} \ , \tag{60}$$

with $z^{'}(x)$ and $z^{''}(x)$ given by (61). Plugging (61) in (60), after some elaborations we get that (60) is negative if and only if

$$\begin{aligned} &4g(g+\alpha) - \alpha\sqrt{(\omega(1+x) + (\alpha+2g))^2 - 4g(g+\alpha)} \\ &\quad < \omega(\alpha+2g)(1+x) + (\alpha+2g)^2 \\ &\quad + \frac{(1+x)^2 x\omega^2 4g(g+\alpha)}{2((\omega(1+x) + \alpha + 2g)^2 - 4g(g+\alpha))}\end{aligned} \tag{62}$$

Let us distinguish two cases. If $\alpha \geq 0$ the left-hand side (LHS) is the sum of a constant and a negative function. On the other hand, the RHS of (62) is the sum of a line with positive slope and of a non-negative function for $x \geq 0$. As a consequence, if $\alpha \geq 0$ (62) holds for any $x \geq 0$ provided that the intercept of the line in the RHS is larger than the constant terms in the LHS. To show this let us evaluate

$$\omega(\alpha+2g) + (\alpha+2g)^2 = \omega(\alpha+2g) + \alpha^2 + 4g^2 + 4\alpha g \ , \tag{63}$$

which is larger than $4g(g+\alpha)$.

Instead, if $\alpha \leq 0$, the LHS is the sum of a constant and a positive function. In this case, a sufficient condition for (62) to hold is

$$\begin{aligned} &4g(g+\alpha) - \alpha\sqrt{(\omega(1+x) + (\alpha+2g))^2 - 4g(g+\alpha)} \\ &\quad < \omega(\alpha+2g)(1+x) + (\alpha+2g)^2 \ ,\end{aligned} \tag{64}$$

wherein we have neglected the positive fraction at the RHS of (62). To show (64), we first show that for $x = 0$ the LHS is smaller than the RHS, and then we show that the LHS grows more slowly than the RHS. For $x = 0$, (64) becomes

$$\begin{aligned} &-\alpha\sqrt{(\omega + \alpha + 2g)^2 - 4g(g+\alpha)} \\ &< \omega(\alpha+2g) + (\alpha+2g)^2 - 4g(g+\alpha) = \omega(\alpha+2g) + \alpha^2 \ .\end{aligned} \tag{65}$$

Since $\alpha \leq 0$, we can square both sides of (65) without changing the direction of the inequality. Then, elaborating we have

$$\begin{aligned} &\alpha^2\left[\omega^2 + \alpha^2 + 2\alpha\omega + 4g\omega\right] \\ &\quad < \alpha^4 + \omega^2\alpha^2 + 4\omega^2 g^2 + 4\omega^2\alpha g + 2\alpha^3\omega + 4\alpha^2\omega g \Leftrightarrow \\ &0 < g + \alpha = \frac{c}{g} \ .\end{aligned}$$

Let us now turn our attention to the derivatives of the LHS and RHS of (64). The first-oder derivative of the LHS is equal to $-\alpha z'(x)$, and is a decreasing function of $x$ because $-\alpha \geq 0$ and $z^{''}(x)$ is a negative function. Then, $z^{'}(x)$ achieves its maximum value for $x = 0$. Instead, the RHS has a positive and constant derivative equal to $\omega(\alpha+2g) = \omega(c/g + g)$. Then, we can argue that the LHS grows more slowly than the RHS provided we can show that

$$-\alpha z'(0) < \omega(\alpha + 2g) \ . \tag{66}$$

Elaborating, (66) is equivalent to

$$\alpha^2\left[1 + \frac{4g(g+\alpha)}{(\omega+\alpha+2g)^2 - 4g(g+\alpha)}\right] < (\alpha+2g)^2 \ , \tag{67}$$

where we have squared both sides of the inequality without changing the direction of the inequality because $-\alpha \geq 0$. Further elaborating reveals that (67) is equivalent to

$$0 < (g+\alpha)(\omega + 4g + 2\alpha) = \frac{c}{g}\left(\omega + 2g + \frac{2c}{g}\right) \ . \tag{68}$$

## Appendix C
## Proof of Proposition

The approach in [6, Theorem 1] can not be directly used due to the fractional nature of (41a). However, this difficulty can be circumvented recalling from Appendix A that a fractional Problem as (41) is equivalent to the problem

$$\max_{\boldsymbol{Q}\succeq 0} \left\{ \log\left|\boldsymbol{I}_{N_B} + \boldsymbol{HQH}^H\right| - \log\left|\boldsymbol{I}_{N_E} + \boldsymbol{GQG}^H\right| - \beta^*(\mu \mathrm{tr}(\boldsymbol{Q}) + P_c)\right\} \tag{69a}$$

$$\text{s.t. } \mathrm{tr}(\boldsymbol{Q}) \leq P_{max} \ , \tag{69b}$$

for the specific $\beta^* \geq 0$ which results in the value of (69a) to be zero. We also recall from Appendix A that the optimal $\beta^*$ is equal to the SEE. Therefore, we will assume $\beta^* > 0$, as the case $\beta^* = 0$ would imply a zero SEE.

Then, we can equivalently work on the KKT conditions of (69). Setting to zero the gradient of the Lagrangian we obtain

$$-(\boldsymbol{I}_{N_A}+\boldsymbol{H}^H\boldsymbol{H}\boldsymbol{Q})^{-1}\boldsymbol{H}^H\boldsymbol{H}+(\boldsymbol{I}_{N_A}+\boldsymbol{G}^H\boldsymbol{G}\boldsymbol{Q})^{-1}\boldsymbol{G}^H\boldsymbol{G} + (\beta^*\mu+\lambda)\boldsymbol{I}_{N_A}-\boldsymbol{M}=0\ ,$$

with $\lambda$ the Lagrange multiplier associated to the maximum power constraint and $\boldsymbol{M}$ the matrix Lagrange multiplier associated to the positive semi-definiteness constraint. Next, multiplying from left by $\boldsymbol{I}_{N_A}+\boldsymbol{H}^H\boldsymbol{H}\boldsymbol{Q}$ and from right by $\boldsymbol{I}_{N_A}+\boldsymbol{Q}\boldsymbol{G}^H\boldsymbol{G}$, and exploring the complementary slackness condition $\boldsymbol{M}\boldsymbol{Q}=\boldsymbol{0}$, we obtain

$$\begin{aligned}&(\beta^*\mu+\lambda)(\boldsymbol{I}_{N_A}+\boldsymbol{H}^H\boldsymbol{H}\boldsymbol{Q})(\boldsymbol{I}_{N_A}+\boldsymbol{Q}\boldsymbol{G}^H\boldsymbol{G})\\&=\boldsymbol{M}+\boldsymbol{H}^H\boldsymbol{H}-\boldsymbol{G}^H\boldsymbol{G}\ .\end{aligned}\tag{70}$$

Next, we follow the argument in [6, Theorem 1] to show that $(\boldsymbol{I}_{N_A}+\boldsymbol{H}^H\boldsymbol{H}\boldsymbol{Q})(\boldsymbol{I}_{N_A}+\boldsymbol{Q}\boldsymbol{G}^H\boldsymbol{G})$ is positive definite. This implies that $(\boldsymbol{M}+\boldsymbol{H}^H\boldsymbol{H}-\boldsymbol{G}^H\boldsymbol{G})\succeq\boldsymbol{0}$, because $\lambda\geq 0$, $\mu>0$, and $\beta^*>0$. Then, at the optimum we have

$$\begin{aligned}0&<\boldsymbol{u}_{+,Q}(\boldsymbol{M}+\boldsymbol{H}^H\boldsymbol{H}-\boldsymbol{G}^H\boldsymbol{G})\boldsymbol{u}_{+,Q}\\&=\boldsymbol{u}_{+,Q}(\boldsymbol{H}^H\boldsymbol{H}-\boldsymbol{G}^H\boldsymbol{G})\boldsymbol{u}_{+,Q}\ ,\end{aligned}\tag{71}$$

where have again exploited the condition $\boldsymbol{M}\boldsymbol{Q}=\boldsymbol{0}$.

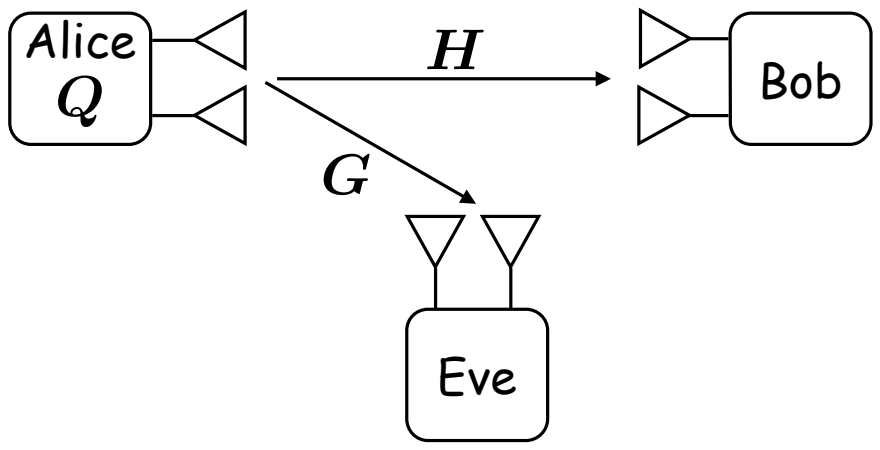

Fig. 1. Considered MIMO-ME system model

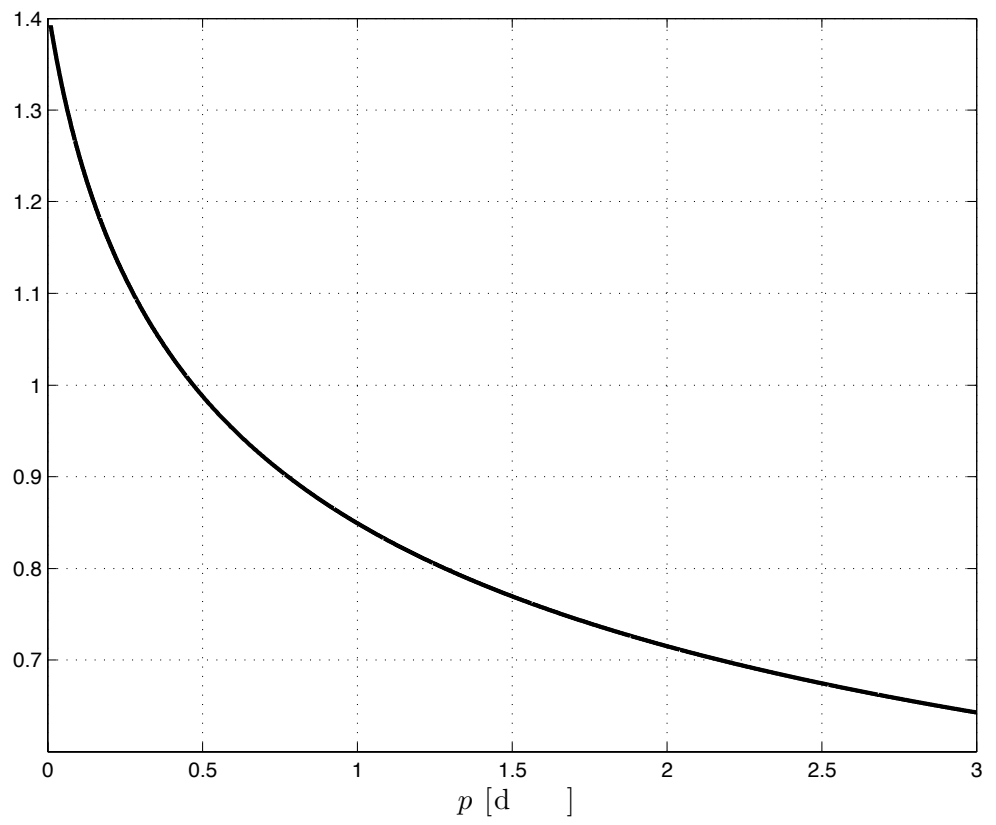

Fig. 2. Function $y(p)$ in (32).

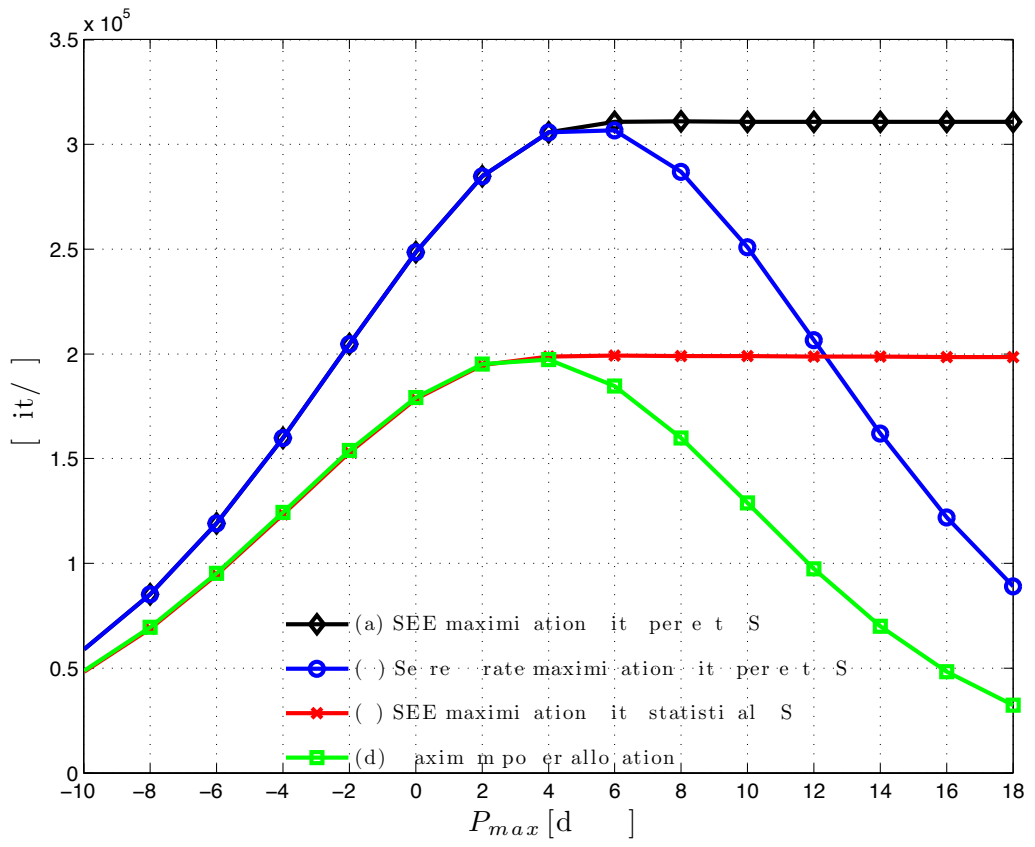

Fig. 3. $N_A=3$, $P_c=5$W. Achieved SEE versus $P_{max}$ for: (a) SEE maximization with perfect CSI; (b) Secrecy rate maximization with perfect CSI; (c) SEE maximization with statistical CSI; (d) Maximum power transmission.

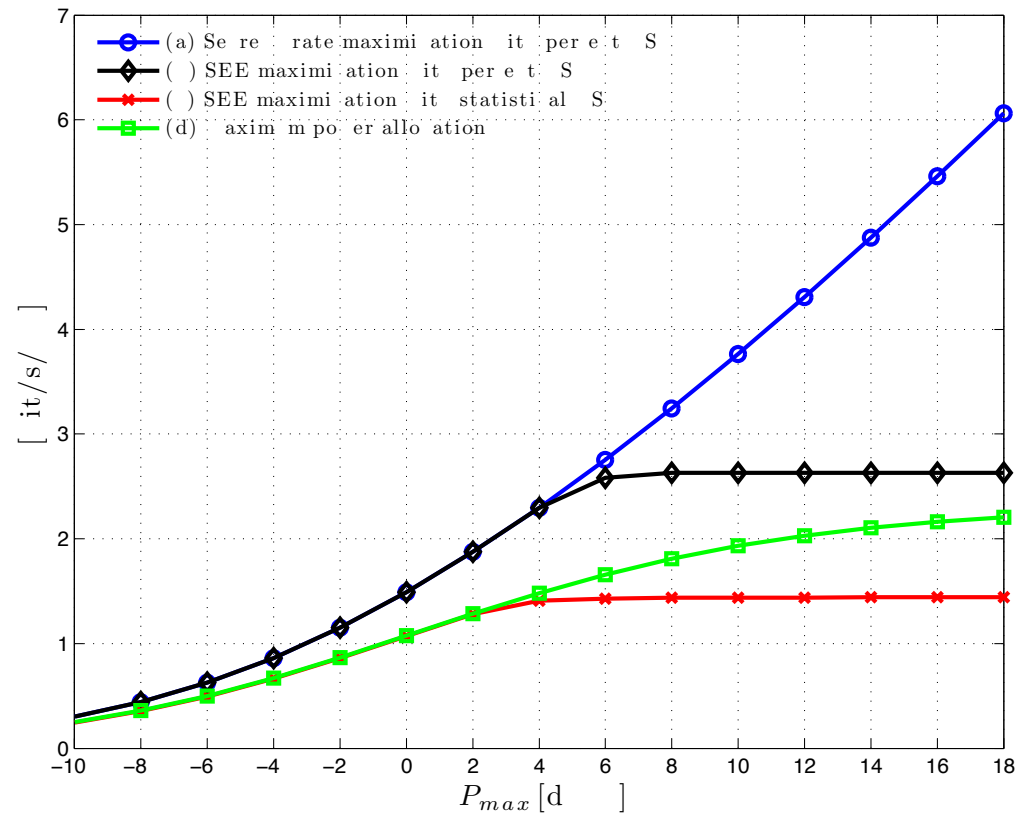

Fig. 4. $N_A=3$. Achieved secrecy rate versus $P_{max}$ for: (a) Secrecy rate maximization with perfect CSI; (b) SEE maximization with perfect CSI; (c) SEE maximization with statistical CSI; (d) Maximum power transmission.

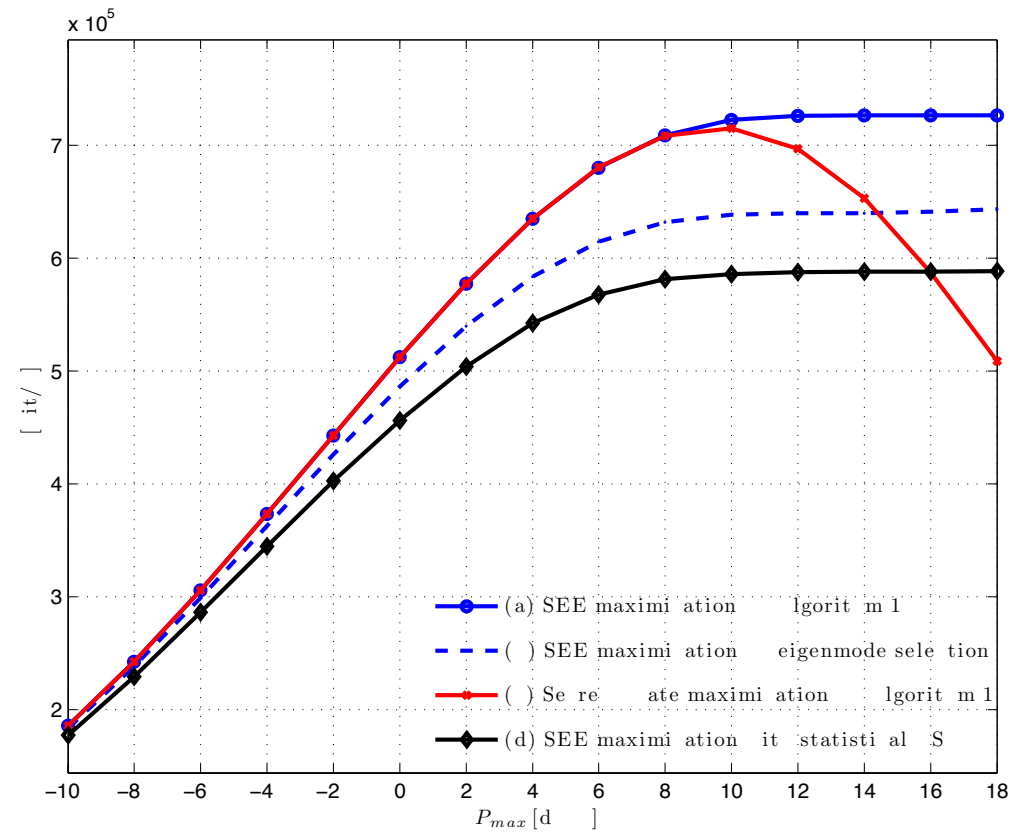

Fig. 5. $N_A=N_B=N_E=2$, $P_c=5$W, $\mu=1$. Achieved SEE versus $P_{max}$ for: (a) SEE maximization with perfect CSI by Algorithm 1; (b) SEE maximization with perfect CSI by eigen-mode selection; (c) Secrecy rate maximization with perfect CSI by Algorithm 1; (d) SEE maximization with statistical CSI by 2.

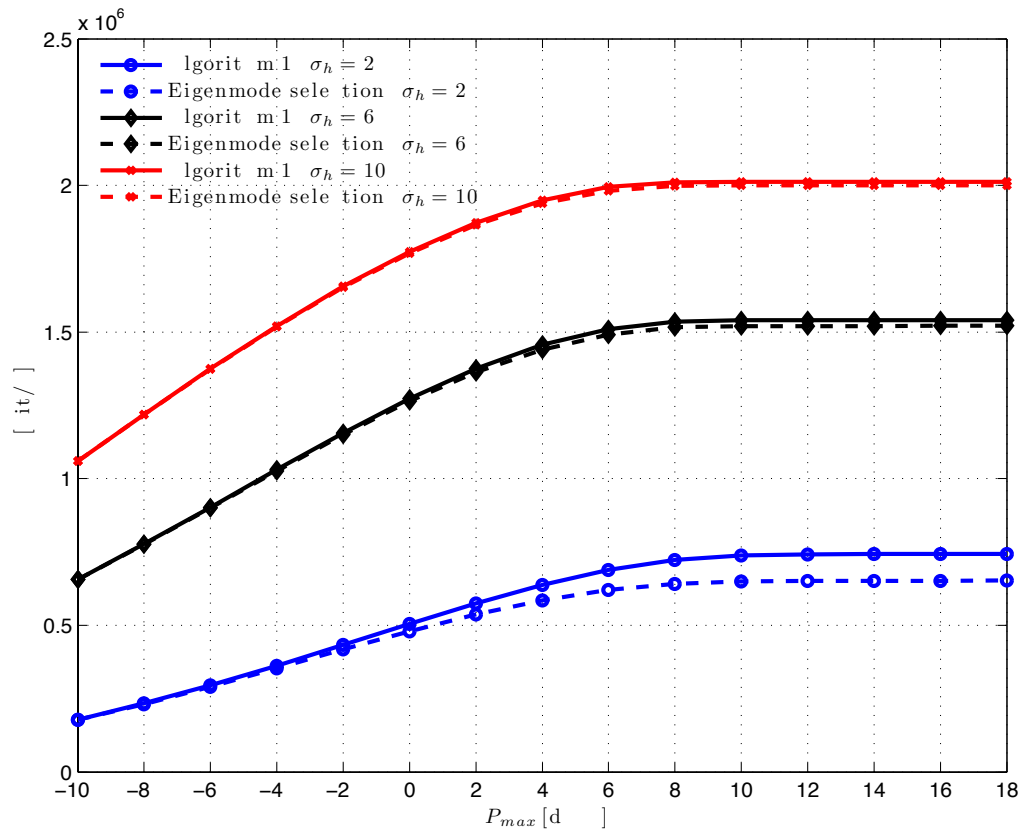

Fig. 6. $N_A = N_R = N_E = 2$, $P_c = 5$W. Achieved SEE versus $P_{max}$ by 1 and Loyka's method, for different values of $\sigma_H$

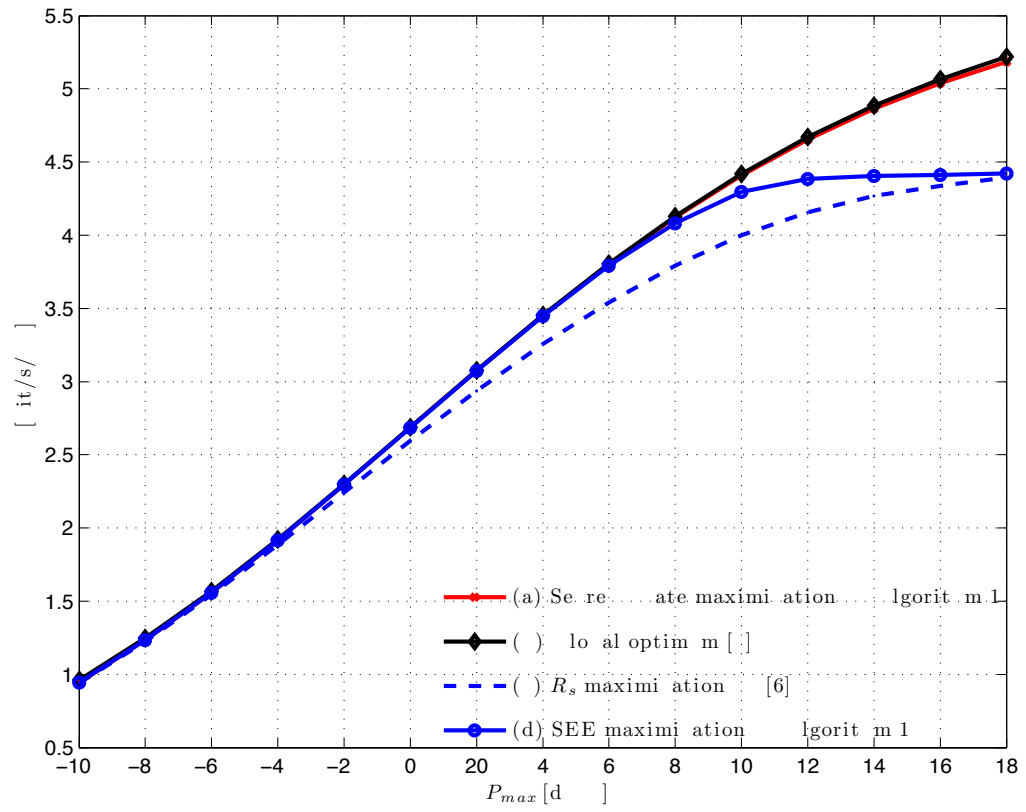

Fig. 7. $N_A = N_B = N_E = 2$, $P_c = 5$ W, $\mu = 1$. Achieved secrecy rate versus $P_{max}$ for: (a) Secrecy rate maximization by Algorithm 1; (b) Secrecy rate global optimum from [8]; (c) Secrecy rate maximization by [6]; (d) SEE maximization by Algorithm 1.